\documentclass{elsart}

\usepackage{enumerate,amsmath,amssymb,mathrsfs}

\newcommand{\W}{\textsc{W}}
\newcommand{\WSAT}{\textsc{W[SAT]}}

\newcommand{\FPT}{\textsc{FPT}}
\newtheorem{observation}{Observation} 

\begin{document}

\begin{frontmatter}

\title{Parameterized Proof Complexity and $\W[1]$}
\author{Barnaby Martin\thanksref{epsrc}}

\address{
School of Engineering and Computing Sciences, Durham University,\\
  Science Labs, South Road, Durham DH1 3LE, U.K.}

\thanks[epsrc]{Supported by EPSRC grant EP/G020604/1.}

\begin{abstract}
We initiate a program of parameterized proof complexity that aims to provide evidence that $\FPT$ is different from $\W[1]$. A similar program already exists for the classes $\W[2]$ and $\WSAT$. We contrast these programs and prove upper and lower bounds for $\W[1]$-parameterized Resolution.
\end{abstract}

\end{frontmatter}

\section{Introduction}

In a series of papers \cite{FOCS2007,FOCS2007journal,Galesietal,BGL-SAT,BGLR} a program of \emph{parameterized proof complexity} is initiated and various lower bounds and classifications are extracted. In \cite{FOCS2007}, parameterized proof complexity is given as a program to gain evidence that $\W[2]$ is different from $\FPT$, while in \cite{FOCS2007journal} the program is recast as searching for evidence that $\WSAT$ is different from $\FPT$. The reason for the discrepancy between the conference and journal versions of that paper is that the latter formulation allows for a cleaner model-theoretic interpretation of the gap theorem which is the paper's principal mathematical result.

Several parameterized proof systems are discussed in \cite{FOCS2007,FOCS2007journal}. The subject of half of the paper is \emph{parameterized tree-Resolution} which is proved to not be \emph{fpt-bounded} in the process of developing a gap theorem for the propositional translations of fo-contradictions in the manner of Riis's well-known gap theorem \cite{complexity_gap}. The system of \emph{parameterized Resolution} is mentioned in the last part of that paper, together with methods to embed parameterized proof systems in classical proof systems. These parameterized proof systems are designed to refute \emph{parameterized contradictions} $(\mathcal{F},k)$, \mbox{i.e.} propositional formulae $\mathcal{F}$ with no satisfying assignment of weight \textbf{$\leq k$}. No lower bounds were given for parameterized Resolution in \cite{FOCS2007,FOCS2007journal} and the first published lower bounds for this system appeared for the \emph{pigeonhole principle} in \cite{Galesietal}. However, the non-fpt-boundedness of parameterized Resolution (and a fortiori of parameterized tree-Resolution) is in fact a trivial observation as the parameterized contradictions
\[ (\Theta_{m,k},k) := \ \ (v^1_1 \vee \ldots \vee v^m_1) \wedge \ldots \wedge (v^1_{k+1} \vee \ldots \vee v^m_{k+1}) \]
are readily seen to require size $\geq m^{k+1}$ to refute (see the forthcoming Observation~\ref{obs:1}).\footnote{This example $(\Theta_{m,k},k)$ seems to have originated independently with Neil Thapen and the present author. It then appeared variously in private communications and it is explained and attributed explicitly to an anonymous referee in \cite{BGLR}.} The most interesting of the further systems discussed in \cite{FOCS2007,FOCS2007journal} we will call \emph{embedded $\W[2]$-parameterized Resolution}. It involves the adding of pigeonhole axioms to directly axiomatise the fact that no more than $k$ of the propositional variables may be evaluated to true. No lower bounds are currently known for this system (in particular the $(\Theta_{m,k},k)$ admit fpt-bounded refutations in this system).

In this note we explore the possibility of a program of parameterized proof complexity that gains direct evidence that $\W[1]$ is different from $\FPT$ (note that the program of \cite{FOCS2007,FOCS2007journal} could be recast anywhere between $\W[2]$ and $\WSAT$, but not lower in the W-hierarchy). Of course, the separation of $\W[1]$ from $\FPT$ is harder than $\W[2]$ from $\FPT$, but we might still wish to attack it directly. Here we seek to refute \emph{weighted parameterized contradictions in $3$-CNF} \mbox{i.e.} propositional $3$-CNF formulae $\mathcal{F}$ with no satisfying assignment of weight \textbf{$= k$}. We go on to explore lower and upper bounds. The trivial lower bound of $(\Theta_{m,k},k)$ no longer applies, and we are forced to look for an alternative. However, the alternative we provide -- although simple -- also gives a lower bound for the new version of \emph{embedded $\W[1]$-parameterized Resolution}.

Another interesting thing about our \emph{$\W[1]$-parameterized Resolution} is that all weighted parameterized contradictions $(\mathcal{F},k)$ created from actual contradictions (\mbox{i.e.} in which $\mathcal{F}$ is a contradiction) become fpt-bounded. This contrasts sharply with the case for parameterized proof systems for $\W[2]$ and above. Indeed, the authors of \cite{BGLR} even suggest to restrict attention to so-called ``strong parameterized contradictions'' $(\mathcal{F},k)$ in which $\mathcal{F}$ is a contradiction itself -- though this is largely in response to the problems posed by $(\Theta_{m,k},k)$. Owing to this, the gap theorem of \cite{FOCS2007,FOCS2007journal} does not fit naturally in this framework, although it can be forced in -- rather as it is in \cite{FOCS2007journal}.

Aside from $(\Theta_{m,k},k)$, the state-of-the-art for parameterized contradictions (coming from actual contradictions) involves non-fpt-boundedness of the pigeonhole principle in $\W[2]$-parameterized bounded-depth Frege in \cite{BGLR}. Similar lower bounds had been derived for $\W[2]$-parameterized tree-Resolution in \cite{FOCS2007} and for $\W[2]$-parameterized Resolution in \cite{Galesietal}.

The paper is organised as follows. After preliminaries and known results, we explore $\W[1]$-parameterized Resolution, with lower bounds in Section \ref{sec:3.2} and upper bounds in Section \ref{sec:3.2}. Finally, we explore lower bounds for embedded $\W[1]$-parameterized Resolution in Section \ref{sec:3.3}.

\section{Preliminaries}

A \emph{parameterized language} is a language $L\subseteq \Sigma^* \times \mathbb{N}$; in an instance $(x,k) \in L$, we refer to $k$ as the \emph{parameter}. A parameterized language is \emph{fixed-parameter tractable} (fpt - and in \FPT) if membership in $L$ can be decided in time $f(k).|x|^{O(1)}$ for some computable function $f$. If FPT is the parameterized analog of P, then (at least) an infinite chain of classes vye for the honour to be the analog of NP. The so-called W-hierarchy sit thus: $\FPT \subseteq \W[1] \subseteq \W[2] \subseteq \ldots \subseteq \WSAT$. For more on parameterized complexity and its theory of completeness, we refer the reader to the monographs \cite{DowneyFellows,FlumGrohe}. Recall that the \emph{weight} of an assignment to a propositional formula is the number of variables evaluated to true. Of particular importance to us is the parameterized problem \textsc{Weighted-3CNF-Sat} (resp., \textsc{Bounded-CNF-Sat}, \textsc{Bounded-Sat}) whose input is $(\mathcal{F},k)$ where $\mathcal{F}$ is a formula in $3$-CNF (resp., CNF, unrestricted) and whose yes-instances are those for which there is a satisfying assignment of weight $=k$ (resp., $\leq k$, $\leq k$). \textsc{Weighted-3CNF-Sat}, \textsc{Bounded-CNF-Sat} and \textsc{Bounded-Sat} are complete for the classes $\W[1]$, $\W[2]$ and $\WSAT$.\footnote{For proofs of latter two completeness results see \cite{FOCS2007,FOCS2007journal}, respectively.} Their respective complements (modulo instances that are well-formed formulae) we call \textsc{PCon}, \textsc{PCon-CNF} and \textsc{W-PCon-3CNF}, complete for the classes co-$\W[1]$, co-$\W[2]$ and co-$\WSAT$. For example, \textsc{PCon} is the language of \emph{parameterized contradictions}, $(\mathcal{F},k)$ \mbox{s.t.} $\mathcal{F}$ has no satisfying assignment of weight $\leq k$; and \textsc{W-PCon-3CNF} contains all $(\mathcal{F},k)$ \mbox{s.t.} $\mathcal{F}$ has no satisfying assignment of weight $= k$.

A \emph{proof system} for a parameterized language $L \subseteq \Sigma^* \times \mathbb{N}$ is a poly-time computable function $P:\Sigma^* \rightarrow\Sigma^*\times \mathbb{N}$ \mbox{s.t.} $\mathrm{range}(P)=L$. $P$ is \emph{fpt-bounded} if there exists a computable function $f$ so that each $(x,k)\in L$ has a proof of size at most $f(k).|x|^{O(1)}$. 
These definitions come from \cite{Galesietal,BGL-SAT,BGLR} and are slightly different from those in \cite{FOCS2007,FOCS2007journal} (they are less unwieldy and have essentially the same properties). The program of \emph{parameterized proof complexity} is an analog of that of Cook-Reckow \cite{Proof_Complexity_start}, in which one seeks to prove results of the form co-$\W[x]\neq$co-$\W[x]$ by proving that parameterized proof systems are not fpt-bounded. This comes from the observation that there is an fpt-bounded parameterized proof system for a co-$\W[x]$-complete $L$ iff $\W[x]=$co-$\W[x]$.

\emph{Resolution} is a refutation system for contradictions $\Phi$ in CNF. It operates on clauses, by the \emph{resolution} rule in which from $(P \vee x)$ and $(Q \vee \neg x)$ one can derive $(P \vee Q)$ ($P$ and $Q$ are disjunctions of literals), with the goal being to derive the empty clause. The only other permitted rule in weakening -- from $P$ to derive $P \vee l$ for a literal $l$. We may consider a Resolution refutation to be a DAG whose sources are labelled by initial clauses, whose unique sink is labelled by the empty clause, and whose internal nodes are labelled by derived clauses. As we are not interested in polynomial factors, we will consider the \emph{size} of a Resolution refutation to be the size of this DAG. Further, we will measure this size of the DAG in terms of the number of variables in the clauses to be resolved -- we will never consider CNFs with number of clauses superpolynomial in the number of variables. 
We define the restriction of Resolution, \emph{tree-Resolution}, in which we insist the DAG be a tree.

The version of parameterized Resolution given for $\WSAT$ in \cite{FOCS2007journal} is a bit awkward in that involves converting non-CNFs to CNF, so we will stick to that in \cite{FOCS2007}. The system of \emph{$\W[2]$-parameterized Resolution} seeks to refute the parameterized CNF contradictions of \textsc{PCon-CNF}. Given $(\mathcal{F},k)$, where $\mathcal{F}$ is a CNF in variables $x_1,\ldots,x_n$, it does this by providing a Resolution refutation of 
\begin{equation}
\mathcal{F}\cup \{\neg x_{i_1}\vee \ldots \vee \neg x_{i_{k+1}} : 1 \leq i_1 < \ldots < i_{k+1} \leq n \}.
\label{equ:W[2]}
\end{equation}
Thus, in $\W[2]$-parameterized Resolution we have built-in access to these additional clauses of the form $\neg x_{i_1}\vee \ldots \vee \neg x_{i_{k+1}}$, but we only count those that appear in the refutation.

In \emph{$\W[1]$-parameterized Resolution}, we seek to refute the weighted parameterized $3$-CNF contradictions of \textsc{W-PCon-3CNF}. Given $(\mathcal{F},k)$, where $\mathcal{F}$ is a $3$-CNF in variables $x_1,\ldots,x_n$, we do this by providing a Resolution refutation of 
\begin{equation}
\begin{array}{ll}
\mathcal{F} & \cup \{\neg x_{i_1}\vee \ldots \vee \neg x_{i_{k+1}} : 1 \leq i_1 < \ldots < i_{k+1} \leq n \} \\
& \cup \{ x_{i_1}\vee \ldots \vee x_{i_{n-k+1}} : 1 \leq i_1 < \ldots < i_{n-k+1} \leq n \}.
\end{array}
\label{equ:W[1]}
\end{equation}


Note that we may consider any refutation system as a $\W[2]$- or $\W[1]$-param- eterized refutation system, by the addition of the clauses given in (\ref{equ:W[2]}) or (\ref{equ:W[1]}), respectively.

Let $[n]:=\{1,\ldots,n\}$. The \emph{pigeonhole principle} will play a role in the paper. Its negation, PHP$_{n+1,n}$, is a contradiction most easily given by the clauses $\neg p_{i,j} \vee \neg p_{l,j}$ ($i\neq l \in [n+1]$ and $j \in [n]$) and $\bigvee_{\lambda \in [n]} p_{i,\lambda}$ ($i \in [n+1]$). PHP$_{n+1,n}$, and its variants, provide contradictions that are ubiquitous in proof complexity, especially since Haken proved an exponential lower bound for it in Resolution \cite{Haken's_classical}.

\subsection{Known Results}

There is a canonical way to translate first-order (fo) sentences $\phi$ to sequences of CNF formulae $\langle\Phi_n\rangle_{n \in \mathbb{N}}$ such that $\Phi_n$ is satisfiable iff $\phi$ had a model of size $n$ (see \cite{FOCS2007} based on \cite{complexity_gap}). When $\phi$ has no finite models, then $\langle\Phi_n\rangle_{n \in \mathbb{N}}$ is a sequence of contradictions ripe for a refutation system. Note that the size of $\Phi_n$ is polynomial in $n$. The famous theorem of Riis goes as follows.
\begin{thm}[Riis 2001 \cite{complexity_gap}]
For an fo-contradiction $\phi$, either: 1.) $\langle\Phi_n\rangle_{n \in \mathbb{N}}$ is refutable in tree-Resolution in size $n^{O(1)}$, or 2.) exists $\epsilon>0$ \mbox{s.t.} every tree-Resolution refutation of $\langle\Phi_n\rangle_{n \in \mathbb{N}}$ is of size $>2^{\epsilon n}$. Furthermore, Case 2 prevails precisely when $\phi$ has some infinite model.
\end{thm}
\noindent In the parameterized setting one might hope for a finer classification of Case 2, and indeed that is what was given in \cite{FOCS2007}.
\begin{thm}[Dantchev, Martin and Szeider \cite{FOCS2007}]
For an fo-contradiction $\phi$ with an infinite model, either: 2a.) $\langle(\Phi_n,k)\rangle_{n \in \mathbb{N}}$ is refutable in $\W[2]$-paramet- erized tree-Resolution in size $\beta^k n^{\alpha}$ ($\alpha$ and $\beta$ constants depending on $\phi$ only), or 2b.) exists $\gamma>0$ \mbox{s.t.} every $\W[2]$-parameterized tree-Resolution refutation of $\langle(\Phi_n,k) \rangle_{n \in \mathbb{N}}$, for $n>k$, is of size $>n^{k^\gamma}$. Furthermore, Case 2b prevails precisely when $\phi$ has a model without a certain kind of finite dominating set.
\end{thm}
\noindent Since Case 2b can be attained, for example when $\phi$ expresses the negation of the pigeonhole principle, it follows that $\W[2]$-parameterized tree-Resolution is not fpt-bounded. The same principle in fact yields a sequence that proves that $\W[2]$-parameterized Resolution is not fpt-bounded \cite{Galesietal}, though this latter fact is somehow trivial in light of the following. Recall $(\Theta_{m,k},k)$ involving $n:=m.(k+1)$ variables.
\begin{observation}
\label{obs:1}
$(\Theta_{m,k},k)$ requires size $\geq m^{k+1}$ (\mbox{i.e.} $\geq \left( \frac{n}{k+1}\right)^{k+1}$) to be refuted in $\W[2]$-parameterized Resolution.
\end{observation}
\noindent This is because we need all $m^{k+1}$ clauses of the form $\neg v^{a_1}_1 \vee \ldots \vee \neg v^{a_{k+1}}_{k+1}$, $a_1,\ldots,a_{k+1} \in \{1,\ldots,m\}$ (comprising the set $\Gamma$), for $\Theta_{m,k} \cup \Gamma$ to be a logical contradiction. That is, for $\Gamma'$ a strict subset of $\Gamma$, $\Theta_{m,k} \cup \Gamma'$ is satisfiable.

If we considered $\W[2]$-parameterized depth-2 Frege (see \cite{Krajicek's_Book} for additional definitions), then $(\Theta_{m,k},k)$ would technically remain not fpt-bounded, because of the large number of additional axioms needed from (\ref{equ:W[2]}). But, those $n^{k+1}$ additional axioms (recall $n:=m(k+1)$) could be coded as a single axiom in depth $2$ -- thus $(\Theta_{m,k},k)$ would essentially become easy. One might reasonably complain, however, that the size of that axiom would be $n^{k+1}$.

\section{Properties of $\W[1]$-parameterized Resolution}
\label{sec:3}

\subsection{Lower bounds}
\label{sec:3.1}

Since $(\Theta_{m,k},k)$ is not a $3$-CNF, it does not provide a trivial reason for $\W[1]$-parameterized Resolution to not be fpt-bounded. However, there is a canonical way to turn CNFs to $3$-CNFs by the use of extension variables, where each clause $(v^1_i \vee \ldots \vee v^m_i)$ becomes, e.g., the conjunction of $(v^1_i \vee v^2_i \vee \neg z^1_i)$, $(z^1_i \vee v^3_i \vee \neg z^2_i)$, \ldots, $(z^{m-4}_i \vee v^{m-2}_i \vee \neg z^{m-3}_i)$, $(z^{m-3}_i \vee v^{m-1}_i \vee v^m_i)$ (where $z^1_i,\ldots,z^{m-3}_i$ are new variables). This or variants thereof is the standard method to reduce SAT to $3$-SAT, known as standard nondeterministic extension in  \cite{AlekhnovichBRW02}. Suppose we thus transform the CNF $(\Theta_{m,k},k)$ to the $3$-CNF $(\Theta'_{m,k},k)$. It is easy to see by the method of a Boolean decision tree solving the search problem, that $(\Theta'_{m,k},k)$ admits $\W[1]$-parameterized \textbf{tree}-Resolution of size $\leq 7^{k+1}$. Of course, this is due to the extra variables we have added and the way in which they contribute to the weight.

Therefore, we must look elsewhere: for $n$ even, define
\[ \Psi_{n,k}:= \ \ \{v_1\leftrightarrow v_2, v_3 \leftrightarrow v_4, \ldots , v_{n-1} \leftrightarrow v_n \}, \]  
where $v_i \leftrightarrow v_{i+1}$ abbreviates $(v_i \vee \neg v_{i+1}), (v_{i+1}\vee \neg v_i)$. Note that $(\Psi_{n,k},k)$ is not a parameterized contradiction, but is a weighted parameterized contradiction.
\begin{lem}
For $k$ odd, $(\Psi_{n,k},k)$ requires size $\geq \left(\frac{n}{2}\right)^{k/2}$ to be refuted in $\W[1]$-parameterized Resolution.
\end{lem}
\begin{pf}
The argument is similar to that for Observation~\ref{obs:1}. We need at least $\left(\frac{n}{2}\right)^{(k+1)/2}$ clauses of the form 
\[\neg v_{2a_1-1} \vee \neg v_{2a_1} \vee \ldots \ldots \vee \neg v_{2a_{(k+1)/2}-1} \vee \neg v_{2a_{(k+1)/2}},\]
(comprising the set $\Gamma$), for $\Psi_{n,k} \cup \Gamma$ to be a logical contradiction. That is, for $\Gamma'$ a strict subset of $\Gamma$, $\Psi_{n,k} \cup \Gamma'$ is satisfiable.
\end{pf}

\subsection{Upper bounds}
\label{sec:3.2}

In the world of $\W[1]$-parameterized Resolution, we have the agreeable situation that non-parameterized contradictions, in $3$-CNF, are always easy to refute (something that does not happen for the $\W[2]$ or $\WSAT$ versions).
\begin{lem}
\label{lem:upper-bound}
Let $\Phi$ be a contradiction in $3$-CNF and let $k$ be arbitrary) then $(\Phi,k)$ has fpt-bounded refutations (of size $\leq 3^{k+1}$) in $\W[1]$-parameterized tree-Resolution.
\end{lem}
\begin{pf}
Prove that $\Phi$ has no satisfying assignment of weight $\leq k+1$ by recursive branching over positive clauses. The positive clauses will not be used up in this process because if they were, this would witness a satisfying assignment for $\Phi$.
\end{pf}
\noindent This means that for lower bounds, one must look at \emph{proper} parameterized contradictions, that \textbf{do} have some satisfying assignments, just not of weight $k$. Our terming of this ``agreeable'' is in sharp contrast to the situation for $\W[2]$, in which the authors of \cite{BGLR} suggest -- as mentioned before -- to restrict attention purely to parameterized contradictions derived from actual contradictions.


\subsection{Embedding into Resolution}
\label{sec:3.3}

We may consider any $3$-CNF weighted parameterized contradiction augmented with pigeonhole clauses enforcing the condition that precisely $k$ variables may be evaluated to true. In this manner, we obtain the system of \emph{embedded $\W[1]$-parameterized Resolution}. A similar system -- \emph{embedded $\W[2]$-parameterized Resolution} -- was presented in \cite{FOCS2007,FOCS2007journal} for parameterized contradictions. 

Given a weighted parameterized contradiction $\Phi_n$, in variables $x_1,\ldots,x_{n}$, we construct $\Phi'_n$ with additional variables $r_{x_1,j},\ldots,r_{x_n, j}$ ($j \in [k]$) and $s_{x_1,j},\ldots,s_{x_n,j}$ ($j \in [n-k]$). The clauses of $\Phi'_n$ are those of $\Phi_n$ augmented by the following pigeonhole clauses. 
\[ \mbox{$\neg x_i \vee \bigvee_{l=1}^{k} r_{x_i,l}$ and $\neg x_i \vee \neg
r_{x_i,j} \vee \neg r_{x_{l},j}$ for $i\neq l \in [n]$ and $j \in [k]$.} \]
\[ \mbox{$x_i \vee \bigvee_{l=1}^{n-k} s_{x_i,l}$ and $x_i \vee \neg
s_{x_i,j} \vee \neg s_{x_{l},j}$ for $i\neq l \in [n]$ and $j \in [n-k]$.} \]
$\Phi'_{n}$ is an ordinary contradiction ripe for an ordinary refutation system (it is no concern that $\Phi'_{n}$ itself is not $3$-CNF). 

Let $\Psi'_{n,k}$ be the CNF generated from the $3$-CNF $\Psi_{n,k}$ by this method. Seeing the two asymmetric pigeonhole principles lurking within $\Psi'_{n,k}$, the following will not be a huge surprise.
\begin{lem}
\label{lem:simple}
Let $k$ be an odd number. Then for no function $f$ and positive integer $c$, can $\Psi'_{n,k}$ be refuted in Resolution in size $f(k).n^c$.
\end{lem}
\begin{pf}
We know from \cite{Haken's_classical} that PHP$_{n+1,n}$ is a family of contradictions without polynomial-size Resolution refutations. Consider the family $\mathcal{P}_{n,k}$ of CNF contradictions in variables $c_1,c_2$, $p_{i,j}$ $(i \in [n]; j\in [k])$ and $s_{i,j}$ $(i \in [n]; j\in [n-k])$, with clauses:
\[
\begin{array}{ll}
\neg c_1 \leftrightarrow c_2 \\
\neg c_1 \vee \neg p_{i,j} \vee \neg p_{l,j} & i \neq l \in [n]; j \in [k] \\
\neg c_1 \vee \bigvee_{\lambda \in [k]} p_{i,\lambda} & i \in [n] \\
\neg c_2 \vee \neg q_{i,j} \vee \neg q_{l,j} & i \neq l \in [n]; j \in [n-k] \\
\neg c_2 \vee \bigvee_{\lambda \in [n-k]} q_{i,\lambda} & i \in [n] \\
\end{array}
\]
\noindent This contains twin pigeonhole principles and is at least as hard to refute as the hardest of these two. To see this, one may consider the decision DAG model with forced assignments to $c_1$ (and therefore $c_2$). The so-restricted decision DAG would then give a refutation of PHP$_{n,k}$ ($c_1$ true) and a refutation of PHP$_{n,n-k}$ ($c_1$ false). In fact, there are never polynomial refutations of $\mathcal{P}_{n,k}$ for any fixed $k$ (as PHP$_{n,n/2}$ is known to be hard \cite{Razborov_WPHP}). However, we need only consider $k=1$ and the fact that $\mathcal{P}_{n,k}$ can therefore not be refuted in Resolution in size $f(k).n^c$ (for any $f$ and $c$).

To complete our proof, we will now reduce $\mathcal{P}_{n,k}$ to $\Psi'_{n,k}$ in order to translate lower bounds of the former to the latter. Using the reduction $x_i := \neg c_2$, $r_{x_i,j}:= p_{i,j}$ and $s_{i,j}:=q_{i,j}$, we claim we can derive in Resolution, in a linear number of steps, the clauses of $\Psi'_{n,k}$ from the clauses of $\mathcal{P}_{n,k}$, whereupon the result follows.

To settle the claim, note that all instances of $v_i \leftrightarrow v_{i+1}$ come from $\neg c_1 \leftrightarrow c_2$. $v_i \vee \neg s_{v_i,j} \vee \neg s_{v_{i'},j}$ and $v_i \vee \bigvee_{l=1}^{n-k} s_{v_i,l}$ are immediate. And $\neg v_i \vee \bigvee_{l=1}^{k} r_{v_i,l}$ and $\neg v_i \vee \neg r_{v_i,j} \vee \neg r_{v_{i'},j}$ involve a series of resolutions with $c_1 \vee \neg c_2$.
\end{pf}
\begin{cor}
Embedded $\W[1]$-parameterized Resolution is not fpt-bounded.
\end{cor}
\noindent We conclude this note with the recollection that the fpt-boundedness of embedded $\W[2]$-parameterized Resolution remains unknown.

\end{document}